\documentclass[12pt,tightenlines,eqsecnum,floats,showpacs,nofootinbib,amsmath,amssymb,aps,prd]{revtex4-2}

\usepackage{tikz}
\usepackage{graphicx,verbatim}
\usepackage{amsmath}
\usepackage{amsfonts}
\usepackage{amssymb}
\usepackage{psfrag}
\usepackage{accents}
\usepackage{mathrsfs}
\usepackage{enumitem}
\usepackage{hyperref}

\newcommand*{\scri}{\ensuremath{\mathscr{I}}} 
\newcommand*{\scrip}{\ensuremath{\mathscr{I}^{+}}} 
\newcommand*{\scrim}{\ensuremath{\mathscr{I}^{-}}} 

\usepackage[colorinlistoftodos]{todonotes}
\usepackage{epstopdf}

\def\be{\begin{equation}}
\def\ee{\end{equation}}
\def\ba{\begin{eqnarray}}
\def\ea{\end{eqnarray}}

\newcommand{\pb}[1]{\hbox{\lower0.5ex\hbox{${}_{\leftarrow}$}}\kern-1.9ex{#1}}

\def\h{\hat}
\def\b{\bar}
\def\ub{\underbar}
\def\ul{\underline}
\def\={\,\hat{=}\,}
\def\f{\frac}

\def\Lie{\mathcal{L}}

\def\bM{\bar{M}}
\def\bg{\bar{g}}
\def\bR{\bar{R}}
\def\bC{\bar{C}}

\def\bm{\bar{m}}
\def\bD{\bar{D}}
\def\bq{\bar{q}}

\def\bS{\bar{S}}
\def\bomega{\bar\omega}
\def\hM{\hat{M}}
\def\hg{\hat{g}}
\def\hR{\hat{R}}
\def\hC{\hat{C}}
\def\hK{\hat{K}}
\def\hD{\hat{D}}
\def\hn{\hat{n}}
\def\hq{\hat{q}}
\def\hS{\hat{S}}
\def\hf{\hat{f}}

\def\ko{\mathring{k}}
\def\j{\bar{\jmath}}

\def\G{\mathfrak{G}}
\def\g{\mathfrak{g}}

\def\B{\mathfrak{B}} 
\def\LBMS{\mathfrak{b}} 

\def\qo{\mathring{q}}
\def\hqo{\hat{\mathring{q}}}

\def\qoub{\mathring{\underbar{q}}}

\def\vo{\mathring{v}}

\def\betao{\mathring{\beta}}
\def\chio{\mathring{\chi}}
\def\varpio{\mathring{\varpi}}
\def\psio{\mathring{\psi}}
\def\alphao{\mathring\alpha}
\def\WIHo{\mathring{\WIH}}

\def\hno{\hat{\mathring{n}}}
\def\uo{\mathring{u}}

\def\Do{\mathring{D}}
\def\omegao{\mathring{\omega}}
\def\epsilono{\mathring{\epsilon}}
\def\hepsilono{\hat{\mathring{\epsilon}}}

\def\WIH{\mathfrak{h}}

\def\psio{\mathring{\psi}}

\def\R{\mathcal{R}}

  \newcommand{\eps}{\epsilon}

\usepackage{pgf}
\makeatletter
\newcommand*{\pgfunderleftarrow}{%
  \@ifstar
    {\let\ifpgf@depth\iftrue\mathpalette\@pgfunderleftarrow}
    {\let\ifpgf@depth\iffalse\mathpalette\@pgfunderleftarrow}%
}
\newcommand*{\@pgfunderleftarrow}[2]{%
  #2%
  \edef\pgf@math@fam{\the\fam}%
  \pgfpicture
    \pgfsetbaseline{0pt}
    \pgf@relevantforpicturesizefalse      
    \pgfsetroundcap                       
    \pgfsetarrowsend{to}
    \pgfutil@tempdima=0.28pt%
    \advance\pgfutil@tempdima by.8\pgflinewidth%
    \pgfutil@tempdima-4\pgfutil@tempdima
    \sbox\pgfutil@tempboxa{$\m@th\fam\pgf@math@fam#1#2$}%
    \advance\pgfutil@tempdima-\dp\pgfutil@tempboxa
    \pgfutil@tempdimb\wd\pgfutil@tempboxa
    \pgfpathmoveto{\pgfqpoint{0pt}{\pgfutil@tempdima}}%
    \pgfpathlineto{\pgfqpoint{-\pgfutil@tempdimb}{\pgfutil@tempdima}}%
    \pgfusepath{stroke}
    \ifpgf@depth
      \pgf@relevantforpicturesizetrue
      \pgfpathmoveto{\pgfqpoint{0pt}{-\pgfutil@tempdimb}}%
      \pgfusepath{use as bounding box}%
    \fi
  \endpgfpicture
}
\makeatother

\begin{document}

\title{Null Infinity as a Weakly Isolated Horizon}
 
\author{Abhay Ashtekar}
\email{ashtekar.gravity@gmail.com}
\affiliation{Institute for Gravitation and the Cosmos, Pennsylvania State 
University, University Park, PA 16802, USA,}
\affiliation{Perimeter Institute for Theoretical Physics, 31 Caroline St N, Waterloo, ON N2L 2Y5, Canada}

\author{Simone Speziale}
\email{simone.speziale@cpt.univ-mrs.fr}
\affiliation{Aix Marseille Univ., Univ. de Toulon, CNRS, CPT, UMR 7332, 13288 Marseille, France}

\begin{abstract}

\medskip

\noindent Null infinity, $\scrip$, arises as a boundary of the Penrose conformal completion $(\hM, \hg_{ab})$ of an asymptotically flat physical space-time $(M, g_{ab})$. We first note that $\scrip$ is a weakly isolated horizon (WIH) in $(\hM, \hg_{ab})$, and then show that its familiar geometric properties can be derived from the general WIH framework \cite{afk,abl1,akkl1,akkl2,akrev,boothrev,Gourgoulhon:2005ng,Jaramillo:2011zw}. This seems quite surprising because physics associated with black hole (and cosmological) WIHs $\Delta$ is \emph{very} different from that extracted at $\scrip$. We show that these differences can be directly traced back to the fact that $\scrip$ is a WIH in the conformal completion rather than the physical space-time. In particular, the  BMS group at $\scrip$ stems from the symmetry group of WIHs \cite{akkl1}. In a companion paper \cite{aass2}, we obtain fluxes and charges associated with symmetries associated with $\scrip$ and $\Delta$ using a new Hamiltonian framework. 
The fact that is there is a single mathematical framework underlying $\Delta$ and $\scrip$ paves the way to explore the relation between horizon dynamics in the strong field region and waveforms at infinity. It should  also be useful in the analysis of black hole evaporation in quantum gravity.
\end{abstract}

\maketitle

\tableofcontents

\section{Introduction}
\label{s1}

This is the third in a series of papers whose goal is to investigate the geometry and physics of quasi-local horizons (QLHs) and relate them to structures available at $\scrip$. QLHs have been analyzed extensively in the literature (for reviews, see, e.g., \cite{akrev,boothrev,Gourgoulhon:2005ng,Jaramillo:2011zw}). The first two papers in this series \cite{akkl1,akkl2} revisited Weakly Isolated Horizons (WIHs) $\Delta$,  and perturbed WIHs, that lie in the physical space-time.%
\footnote{Titles of \cite{akkl1,akkl2} referred to Non Expanding Horizons (NEHs) rather than WIHs. However, while the starting point was indeed NEHs, the subsequent discussion was based on the canonical WIH structure that NEHs  naturally inherit.} 
Using a new perspective, these papers discussed the universal structure and symmetry group $\G$ of $\Delta$, as well as the associated charges and fluxes across perturbed $\Delta$. That discussion showed that, even though WIHs in physical space-times lie in strong curvature regions, there is a surprising similarity between $\G$ and the BMS group $\B$ that refers to the asymptotic region, and also between the expressions of fluxes across perturbed WIHs and those across $\scrip$, associated with the respective symmetry generators.

In this paper we will investigate the complementary issue: Recovering the structure at $\scrip$ starting from the WIH framework. We will show that $\scrip$ is naturally endowed with the structure of a WIH --\emph{not} a perturbed WIH but a \emph{proper} WIH-- even when there is a large flux of radiation across it! This seems very surprising at first since WIHs are generally associated with the black hole (or cosmological) horizons $\Delta$ that lie in the strong curvature region and there is no flux of gravitational radiation (or matter fields) across them. Therefore WIHs have had the connotation of representing geometries of boundaries that are in equilibrium; indeed WIHs are often regarded as being `only slightly weaker than Killing horizons'. 

Goal of this paper is to resolve this apparent tension. In a nutshell, we will show that the difference can  be traced back to the fact that, whereas black hole (or cosmological) WIHs $\Delta$ are sub-manifolds of the physical space-time $(M, g_{ab})$, $\scrip$ is not. It is a submanifold of the Penrose completion $(\hM, \hg_{ab})$, and therefore a WIH in $(\hM, \hg_{ab})$. Nonetheless, $\Delta$ and $\scrip$  share a number of key geometric features. In particular, they both inherit from the 4-dimensional space-time metric a preferred family of null normals, a degenerate metric, and an intrinsic connection with properties that endow them with the structure of a WIH. Furthermore, their universal structures and symmetry groups are essentially the same. However, as we will show, because $\Delta$ is a WIH with respect to the physical metric $g_{ab}$, the time dependence of the intrinsic connection $D$ is entirely determined from Einstein's equations by the values of certain fields on a 2-sphere cross-section of $\Delta$. In this sense $D$ does not carry `3-dimensional' degrees of freedom, and the physical information encoded in it is purely `Coulombic'.
By contrast, at $\scrip$ we have \emph{conformal} Einstein's equations and they constrain the intrinsic connection $\hD$ on $\scrip$ only mildly: Now it carries precisely the two  \emph{freely specifiable, radiative} degrees of freedom per point of the 3-manifold $\scrip$. In striking contrast to the connection on $D$ on $\Delta$, the connection $\hD$ on $\scrip$ has no information about the Coulombic aspects of the gravitational field. Thus the mathematical difference between Einstein's equations satisfied by $g_{ab}$ and conformal Einstein's equations satisfied by $\h{g}_{ab}$ translates to a striking difference between the physics they capture! It is rather astonishing that very different physics can emerge from a common WIH framework.

These results are presented in sections \ref{s2} and \ref{s3}. In section~\ref{s2} we begin by recalling the WIH framework in a general space-time $(\bM, \bg_{ab})$ that \emph{does not refer to any field equations} and summarize the structure on a generic WIH $\WIH$. We then discuss the additional structure that arises on the the black hole (or cosmological) WIHs $\Delta$ which are sub-manifolds of the physical space-times $(M, g_{ab})$ satisfying Einstein's equations (possibly a cosmological constant and/or  matter). Finally we show that $\scrip$ is a WIH in the conformal completion $(\hM, \hg_{ab})$ of an asymptotically flat space-time in which the stress-energy tensor has the standard fall-off. It is a rather special WIH: When equipped with the standard null normal $\hn^a$ endowed by the conformal completion, it is a non-rotating, extremal WIH in $(\hM, \hg_{ab})$. The geometrical fields one is accustomed to see on $\scrip$ (summarized, e.g., in \cite{gerochrev,aa-yau}) are all directly induced by its WIH structure \cite{afk,abl1,akkl1}. In section~\ref{s3} we discuss the universal structure and associated symmetries. The symmetry group $\G$ of a generic WIH is a 1-dimensional extension of the BMS group $\B$ \cite{akkl1}. We pinpoint the extra structure that $\scrip$ naturally inherits vis a vis generic WIHs that reduces $\G$ to $\B$. Thus the BMS group can be arrived at from a WIH perspective. In the companion paper \cite{aass2} we introduce a new  phase-space framework tailored to degrees of freedom that ``reside in local 3-d regions''. By specializing this framework to regions of $\scrip$ and $\Delta$ we will obtain the expressions of flux and charge observables associated with the corresponding symmetries. Again, one and the same framework leads to dramatically different predictions: At $\scrip$ we recover the standard BMS charges and fluxes while on  $\Delta$ the fluxes vanish identically, just as one would physically expect. This occurs because of the difference in the nature of degrees of freedom in the two cases which, in turn, arises because while we have conformal Einstein's equations at $\scrip$, we have Einstein's equations at $\Delta$.

Our conventions are as follows.  While discussing the general framework, we denote space-times by $(\bM, \bg_{ab})$ and WIHs by $\WIH$. In the discussion of black hole and cosmological horizons $\Delta$, the underlying 
physical space-times is denoted by $(M, g_{ab})$, and the Penrose conformal completions used in the discussion of $\scrip$ is denoted by  $(\hM, \hg_{ab})$. The torsion-free derivative operator compatible with $\bg_{ab}$ is denoted by $\bar\nabla$ and its curvature tensors are defined via:  $2\bar\nabla_{[a}\bar\nabla_{b]} v_c = \bR_{abc}{}^d v_d$, $\bR_{ac} = \bR_{abc}{}^b$, and $\bR=g^{ab}\bR_{ab}$. All fields are assumed to be smooth for simplicity but this requirement can be weakened substantially (in particular to allow for the possibility that the Newman-Penrose curvature component $\Psi_1^\circ$ may violate peeling). \emph{If there is a possibility of ambiguity, we will use $\=$ to denote equality that holds only at the WIHs.} Finally, null normals of $\WIH$ are assumed to be future directed. In the discussion of null infinity, we will focus on $\scrip$ for definiteness. But it is obvious that all our considerations apply to $\scrim$ as well. 

The main results of this paper are summarized in a brief report~\cite{Ashtekar:2024mme}. After this work was completed and reported there, we learned of another work that discusses the relation between null infinity and horizons from a somewhat different perspective \cite{Freidel:2024tpl}. 

\section{Geometry of WIHs}
\label{s2}
The structure of quasi-local horizons has been discussed extensively in the literature (e.g., \cite{afk,abl1,akrev,boothrev,Gourgoulhon:2005ng,Jaramillo:2011zw}). However, since the primary focus of these works was on black hole and cosmological horizons $\Delta$, these discussions begin by restricting themselves to horizons that lie in the physical space-time. In section~\ref{s2.1} we drop this assumption: the structure we discuss is common to both $\Delta$ and $\scrip$. In section~\ref{s2.2} we discuss the additional structure that becomes available on $\Delta$, and in section~\ref{s2.3}, on $\scrip$.
\subsection{General Framework}
\label{s2.1}
Consider a 4-manifold $\bM$ equipped with a metric $\bg_{ab}$ of signature -,+,+,+. We will begin with a simpler and more general notion of non-expanding horizons and then arrive at WIHs.\vskip0.05cm
\textbf{Definition 1}:  A null 3-dimensional sub-manifold $\WIH$ of $\bM$ will be said to be a \emph{non-expanding horizon} (NEH) if \cite{akkl1}: \\
(i) Its topology is $\mathbb{S}^2\times \mathbb{R}$;\\ 
(ii) Every null normal ${\bar{k}}^a$ to $\WIH$ is expansion free:\,\, $\theta_{({\bar{k}})} \= 0$ where $\=$  \emph{stands for equality at points of\, $\WIH$}. For definiteness, we will assume that the null normals ${\bar{k}}^a$ are all future pointing;\\
(iii) The Ricci tensor $\bR_a{}^b$ of $\bg_{ab}$ satisfies $\bR_a{}^b {\bar{k}}^a\, \= \,\alpha {\bar{k}}^b$ for some function $\alpha$.\vskip0.1cm 
\noindent An NEH can be a proper sub-manifold of $\bM$, or a part of its boundary. Motivation behind the first two conditions is obvious. The last condition (iii) is a restriction only on geometry; it does not refer to matter fields.  It is motivated by the following two considerations:\\
(a) if $\WIH$ were to be an NEH $\Delta$ in the physical space-time in which the 4-metric satisfies Einstein's equation, then (iii) is necessary and sufficient for the local flux  $T_{ab} \xi^a {\bar{k}}^b$ across $\Delta$ to vanish for all vector fields $\xi^a$ that are tangential to $\Delta$ (which, in particular include the symmetry vector fields of $\Delta$ \cite{akkl1}); and, \\
(b) it automatically holds at $\scrip$ for the conformally rescaled metric, if the physical metric satisfies Einstein's equations with stress-energy that has the standard fall-off.\vskip0.05cm

NEHs have several interesting properties that follow directly from the definition:

(1) One can always choose the null normals to $\WIH$ to be affinely parametrized geodesic vector fields ${\bar{k}}^a$. Any two of them are related by ${\bar{k}}^{\prime\,a} \= f\, {\bar{k}}^a$ where $f$ satisfies $\Lie_{{\bar{k}}} f \=0$. Let us restrict ourselves to such null normals.

(2) The pull-back $\bq_{ab}$ of the space-time metric $\bg_{ab}$ to $\WIH$ has signature 0,+,+.  The Raychaudhuri equation, together with condition (iii) of Definition 1, implies that $\bq_{ab}$  satisfies: $\bq_{ab} {\bar{k}}^b \=0$ and $\Lie_{{\bar{k}}}\, \bq_{ab} \= 0$. Thus the shear $\sigma_{ab}^{({\bar{k}})}$ of every null normals also vanishes. 
\footnote{Many of the properties of NEHs continue to hold if condition (iii) is replaced by a weaker requirement $\sigma_{ab}^{({\bar{k}})}\, \hat{=}\, 0$.  However, then the Raychaudhuri equation only implies that $\bR_{ab} {\bar{k}}^a {\bar{k}}^b \,\hat{=}\, 0$ which is insufficient for some of the results, e.g., the characterization of the geometry of black hole and cosmological horizons in terms of multipoles \cite{aepv,akkl2}. In vacuum, NEHs coincide with expansion-free null surfaces (which are automatically shear-free) but are a subset of expansion and shear-free surfaces in the presence of matter and/of a cosmological constant.}

(3) Since the expansion and shear of ${\bar{k}}^a$ vanish on $\WIH$, it follows that the space-time derivative operator $\bar\nabla$ compatible with $\bg_{ab}$ \emph{induces a unique derivative operator $\bD$ on $\WIH$ via pull-back}.
$\bD$ `interacts' with the degenerate metric $\bq_{ab}$ and the null normals ${\bar{k}}^a$ in an interesting fashion. On $\WIH$ we have:
\be \label{eq1} \bD_a  \bq_{bc}\, \= \,0 \quad {\rm and} \quad \bD_a {\bar{k}}^b\, \= \,\bomega_a\,{\bar{k}}^b, \ee
for some 1-form $\bomega_{a}$. Under ${\bar{k}}^a \to {{\bar{k}}}^{\prime\, a} = f {\bar{k}}^a$ (with $\Lie_{{\bar{k}}} f \=0$), we have $\bomega_a \to \bomega^\prime_a = \bomega_a + D_a \ln f$. Thus, $\bomega_a$ depends on the choice of the geodesic null normal ${\bar{k}}^a$. But we will suppress this dependence for notational simplicity. For any choice of ${\bar{k}}^a$, the corresponding $\bomega_a$ also satisfies
\be \label{eq2} \bomega_a{\bar{k}}^a\, \=\,0, \quad {\rm and} \quad \Lie_{{\bar{k}}} \bomega_a \, \=\,0\, . \ee
Thus, $\bq_{ab}$ and $\bomega_a$ are pull-backs to $\Delta$ of covariant tensor fields $\underbar{q}_{ab}$ and $\underline{\bomega}_a$ on the 2-sphere $\underline{\WIH}$ of integral curves of the null normals ${\bar{k}}^a$. 

In the literature on null hypersurfaces one often introduces a `rigging derivative' defined using a choice of auxiliary 1-form, see e.g. \cite{cfp,Chandrasekaran:2021hxc}. This is also referred to as `Carrollian connection' \cite{Ciambelli:2019lap}. It is easy to show that when the hypersurface is shear and expansion-free --as is the case for $\WIH$-- this rigging/Carollian connection becomes independent of the rigging vector and coincides with our $\bD$.

(4) One can essentially exhaust the rescaling freedom in the choice of (the affinely parametrized geodesic null normals) ${\bar{k}}^a$ on $\WIH$ by requiring that $\underline\omega_a$ be divergence-free on $\underline\WIH$. This selects a small equivalence class of null normals, where two are equivalent if they are related by rescaling by a positive \emph{constant}. All ${\bar{k}}^a$ in this equivalence class share the same 1-form $\bomega_a$ that satisfies $\bD^a \bomega_a \equiv \bq^{ab} \bD_a \bomega_b\=0$ on $\WIH$ (where $\bq^{ab}$ is any inverse of $\bq_{ab}$). There is no natural structure on a generic $\WIH$ to eliminate the freedom of constant rescalings. This fact will turn out to be important in section \ref{s3.1}.  We will denote the equivalence class of these preferred geodesic null normals by $[{\bar{k}}^a]$. From now on we will \emph{restrict ourselves to ${\bar{k}}^a$ that belong to this 1-parameter family.}
\medskip

\begin{table}
\begin{center}
\footnotesize
\begin{tabular}{|l|c|c|c|}
\hline
\hspace{1.5cm}\emph{ Notions}\quad   &\,\,\, \emph{General spacetimes}\,\,\, & \,\,\,\emph{Physical spacetimes}\,\,\, &\,\,\, \emph{Conformal Completions}\,\,\,\\ \hline\hline
Field Eqs &  None & Einstein's Eqs & Conf. Einstein's Eqs \\ \hline
4-Manifolds \&  metrics thereon  & $\b{M}, \b{g}_{ab}$ & ${M}, {g}_{ab}$ & $\h{M}, \h{g}_{ab}$ \\ \hline
Horizons & $\WIH$ & $ \Delta^+$ & $\scrip$ \\ \hline
Induced metrics & $\b{q}_{ab}$&$q_{ab}$&$\h{q}_{ab}$ \\ \hline
Null normals &$\b{k}^a$&$\ell^a$&$\hn^a$ \\ \hline
Dual 1-forms & $\bar\jmath_a$ &$n_a$ &$\hat\ell_a$ \\\hline
Intrinsic derivative operators & $ \b{D}$&$D$&$ \hD$\\\hline
Local degrees of freedom& & $(q_{ab}, D_a n_b)|_{\mathbb{S}^2}$ & $\{  \hD  \}$ \\
&& Purely Coulombic & Purely Radiative \\\hline
Universal structure & $\mathring{q}_{ab},\, [\mathring{k}^a]$ & $\mathring{q}_{ab},\, [\mathring{\ell}^a]$ 
& $\hat{\mathring{q}}_{ab},\, \hat{\mathring{n}}^a$\,\,\, or\,\,\, $\h{q}_{ab}, \hn^a$ \\\hline
Symmetry vector fields &$\xi^a$&$\xi^a$&$\xi^a$ \\\hline
Symmetry groups & $\G=\B \ltimes \mathfrak{D}$ &$\G$ & $\B$ \\\hline
\end{tabular}
\caption{\emph{Symbols associated with WIHs and their meaning}}
\label{tab1}
\end{center}
\end{table}

In the terminology used in the literature \cite{afk,abl1,akrev,boothrev,Gourgoulhon:2005ng,Jaramillo:2011zw} a \emph{Weakly Isolated Horizon} is defined as an NEH $\WIH$ equipped with an equivalence class $[\bar{k}^a]$ of null normals satisfying $\Lie_{\b{k}} \b\omega_a\, \=\,0$. (Note that normals $\bar{k}^a$ in this equivalence class need not be affinely parametrized geodesics, i.e. while $\bomega_a \bar{k}^a$ is constant because  $\Lie_{\b{k}} \b\omega_a\, \=\,0$, the constant need not be zero.) On these horizons both $\bq_{ab}$ and the part of the connection $\bD$ encoded in $\bomega_a$ are time-independent. WIHs have been extensively studied in the literature on mechanics of quasi-local horizons (see, e.g., \cite{afk,abl2}) and on characterization of horizon geometries using multipoles (see, e.g., \cite{aepv,akkl1}). 
In our case we did restrict $\bar{k}^a$ to be affinely parametrized geodesic vector case whence
it follows from Eq. (\ref{eq2}) that our choice of null normals $\bar{k}^a$ endows $\WIH$ with the structure of a WIH. Furthermore, since integral curves of any of our null normals $\bar{k}^a$ are affinely parametrized null geodesics --i.e. since $\b\omega_a \b{k}^a \=0$-- the resulting WIH is said to be \emph{extremal}.  Properties (1-4) reviewed above show that any NEH can be naturally endowed with the structure of an extremal WIH. This is the class of the WIHs we will work with. \vskip0.1cm

The \emph{WIH geometry} is encoded in the triplet $(\bq_{ab}, [{\bar{k}}^a], \bD)$. As noted above, $\bq_{ab}$ and $\bomega_a$ are time-independent. However, they do not determine $\bD$ uniquely. In fact, $\bD$ is generically \emph{time-dependent}! Because of this key feature, WIHs cannot be regarded as \emph{stationary} horizons. Let us analyze the time dependence of $\bD$. Properties discussed under points (2) and (3) above imply that if a 1-form $\bar{h}_a$ is `horizontal' --i.e. satisfies $\bar{h}_a \bar{k}^a \=0$, then the action $\bD_a \bar{h}_b$ of $\bD$ is completely determined by $\bq_{ab}$ and is
time independent: $(\Lie_{{\bar{k}}} \bD_a - \bD_a \Lie_{{\bar{k}}}) \bar{h}_b \=0$. 
Thus, the time dependence of $\bD$ is encoded in the action $\bD_a \j_b$ on any 1-form $\j_b$ on $\WIH$ satisfying\, $\j_b {\bar{k}}^b \hat{=} -1$. Now, ${\bar{k}}^b \bD_a \j_b \= \bomega_b$ which is again time-independent. However, the rest of $\bD_a \j_b$ is not. To evaluate this time dependence, let us consider any 1-form $\b{\j}_b$ in a neighborhood of $\WIH$ whose pull-back to $\WIH$ is $\j_b$, and use the fact that $\bD_a \j_b$ is the pull-back to $\WIH$ of $\bar\nabla_a \b\j_b$ to relate curvature of $\bD$ to that of $\bar\nabla$. Then it it follows that the `time derivative' of $\bD$ is given by
\be \dot{\bD}_a \j_b\, :\!\!\=\, (\Lie_{{\bar{k}}} \bD_a - \bD_a \Lie_{{\bar{k}}}) \j_b \,\=\, \bD_a\bomega_b + \bomega_a \bomega_b + {\bar{k}}^c\, \bR_{c\, {\pb{ab}}}{}^d\, \j_d \ee
where, on the right hand side, the indices $a,b$ are pulled back to $\WIH$. Expanding the Riemann tensor of $\bg_{ab}$ in terms of the Weyl and Schouten tensors, $C_{abcd}$ and $\bS_{ab} = \bR_{ab} -\f{1}{6}\bR\,\bg_{ab}$, recalling that $\j_b {\bar{k}}^b \= -1$,  and using condition (iii) of Definition 1, namely, $\bR_{a}{}^{b}{\bar{k}}^a = \alpha {\bar{k}}^b$,  one obtains
\be \label{ddot} \dot{\bD}_a \j_b \= \bD_a\bomega_b + \bomega_a \bomega_b + {\bar{k}}^c\, \bC_{c\,{\pb{ab}}}{}^d\, \j_d + \f{1}{2}\,\big(\bS_{\pb{ab}} + (\alpha\, - \f{1}{6}\b{R})\, \bq_{ab}\big)\, . \ee
On a generic WIH, none of the terms on the right side are zero, whence \emph{$\bD$ has time dependence}. By contrast, on \emph{isolated} horizons (IHs), $\dot{\bD} \=0$ by definition \cite{Ashtekar:2000sz}; IHs are much more restrictive. In particular, as we saw, one can always choose a null normal on an NEH to endow it with the structure of a WIH. However, generically there is no null normal that can endow it with an IH structure. Stationary or Killing horizons are \emph{even more} restrictive than IHs because the full 4 metric $\b{g}_{ab}$ and its derivative operator ${\bar\nabla}$ as well as curvature tensors --not just $\bq_{ab}$ and ${\bD}$-- is now time independent on $\WIH$. 

Finally, let us note an interesting fact that will be useful especially in our analysis of $\scrip$. Suppose we are given a space-time $(\bM, \bg_{ab})$ that admits a WIH $\WIH$. Then, $\WIH$ is also a WIH in a conformally related space-time $(\bM, \bg^\prime_{ab} = \mu^2 \bg_{ab})$ for a smooth, non-zero conformal factor  $\mu$ if and only if $\Lie_{{\bar{k}}} \mu\, \= \,0$ for any null normal ${\bar{k}}^a$ to $\WIH$. Thus, the property that a sub-manifold $\WIH$ is a WIH is shared by a conformal family of 4-metrics, related by a conformal factor that is time independent on $\WIH$. Of course, the two WIH-geometries would be distinct because $(\bq_{ab}^{\prime},\, \bD^\prime) \not= (q_{ab},\, \bD)$.\\

\emph{Remark:}  As noted above, a general WIH is an NEH equipped with equivalence classes $[\b{k}^a]$ of null normals\, --where $\b{k}^{\prime\,a} \approx \b{k}^a$ iff $\b{k}^{\prime\,a} = c \b{k}^a$ for a positive constant $c$-- \,for which $\Lie_{\b{k}} \b\omega_a \=0$. Then it follows that $\b\omega_a \b{k}^a = \kappa$, a constant on $\WIH$ \cite{afk}. WIHs are naturally divided into two classes: the extremal WIHs, considered so far, on which $\kappa =0$, and,  \emph{non-extremal} WIHs on which $\kappa\not= 0$. The main differences between these two cases can be summarized as follows:\\ 
(i) On the extremal WIHs, the acceleration of \emph{any} null normal in the equivalence class $[\b{k}^a]$ is the same, namely zero, while in the non-extremal case if $\b{k}^{\prime\,a} = c \b{k}^a$, then $\kappa^\prime = c \kappa$. In the non-extremal case, the value of $\kappa$ is not a property of the WIH itself but of a specific null normal in the equivalence class one chooses; and,\\ 
(ii) While every NEH $\WIH$ naturally admits a canonical extremal WIH structure, there is no natural procedure to select an equivalence class of null normals $[\b{k}^a]$ that endows the given $\WIH$ with a non-extremal WIH structure.\\ 
Note that whether a given NEH is extremal or not depends on which null normals $[\b{k}^a]$ one chooses to emphasize. For instance, since the Schwarzschild horizon is also a Killing horizon, it is natural to ask that $[\b{k}^a]$ be the restriction to $\WIH$ of a Killing field, and, in the $\Lambda=0$ case, to fix also the rescaling freedom by demanding that the Killing field be unit at infinity. We then have a non-extremal IH.
\footnote{But the null normals $\b{k}^a$ that endow the Schwarzschild horizon with an extremal WIH structure also have interesting applications, e.g., in the definition of the Unruh vacuum in quantum field theory on the Schwarzschild background. By contrast, the standard non-extremal null normal $\b{k}^a$ that descends from the normalized static Killing field corresponds to choosing the Boulware vacuum.}

Being NEHs, the non-extremal WIHs also inherit a canonical derivative operator $\bD$ from space-time whose action is again completely determined by $\bD_a \j_b$. But now Eq.~(\ref{ddot}) providing its time dependence acquires an extra term proportional to $\kappa$ \cite{abl1}. However, the structure of the equation is the same and our main conclusions of section \ref{s2.2} continue to hold also in the non-extremal case. The symmetry group in the non-extremal case is also the same as that in the extremal case discussed in section \ref{s3.1}. 

In this paper the focus is on \emph{extremal} WIHs because: (i) As we saw, one can naturally endow any NEH horizon with a canonical extremal WIH structure, and, (ii) As we will show in \ref{s2.3}, in the conformal completions $(\hM, \hg_{ab})$ commonly used in the literature, $\scrip$ is an \emph{extremal} WIH.
 
\subsection{WIHs $\Delta$ in physical space-times}
\label{s2.2}
Let us now apply the general framework of section~\ref{s2.1} to the black hole and cosmological WIHs. These lie in a physical space-time satisfying Einstein's equations. There is extensive literature on these WIHs (see, e.g., \cite{afk,abl1,akrev,boothrev,Gourgoulhon:2005ng,Jaramillo:2011zw}). In these investigations, physical considerations led to the assumption that the space-time metric satisfies Einstein's equations, possibly with a cosmological constant, with Maxwell and Yang-Mills fields as sources. In this subsection, we will show that this restriction results in certain simplifications and provides some additional structures vis a vis our discussion of section \ref{s2.1}. Interestingly, these in turn imply that there are no radiative degrees of freedom on $\Delta$. This reflects the fact that there is no flux of radiation across these WIHs.

Let us then restrict ourselves to WIHs that are sub-manifolds of the physical space-time $(M, g_{ab})$ which satisfies Einstein's equations (possibly with a cosmological constant and matter sources). To facilitate comparison with the existing literature, in this sub-section we will use notation introduced there.  Let us now denote the WIHs by $\Delta$ (rather than $\WIH$) and the preferred equivalence class of null normals by $[\ell^a]$ (rather than $[{\bar{k}}^a]$) . Given a specific null normal $\ell^a \in [\ell^a]$, the `conjugate' 1-form on the WIH will now be denoted by $n_a$ (rather than $\j_a$), so that $\ell^a n_a \= -1$, and, if $\ell^{\prime\,a}\, \=\, c \ell^a$ for a positive constant $c$, then $n^\prime_a \, \=\, c^{-1} n_a$. There is considerable freedom in the choice of $n_a$ but the main results are insensitive to the specific choice.  As in the existing literature, we will denote the intrinsic derivative operator on $\Delta$ by $D$ (rather than $\bD$), so that the 1-form $\omega_a$ is now given by $D_a \ell^b\, \=\, \omega_a \ell^b$, or $\omega_a\, \= - n_b D_a \ell^b$. Eq.(\ref{eq2}) now implies that $\Delta$ is in fact \emph{an extremal, weakly isolated horizon (WIH)}, i.e., a WIH on which $\Lie_\ell \omega_a \=0$ and surface gravity $\kappa_{(\ell)} := \omega_a \ell^a$ vanishes. 

There are some interesting constraints on the space-time Weyl curvature at $\Delta$. (These are also present on the WIHs\,  $\WIH$\, of section~\ref{s2.1}.) We will now make them explicit by introducing a standard Newman-Penrose null tetrad. Pick an $\ell^a$ in the equivalence class $[\ell^a]$. So far the 1-form $n_a$ is defined intrinsically on $\Delta$. Let us begin by extending it to the 4-d tangent space as a \emph{null} 1-form which, for simplicity of notation, we will again denote by $n_a$. Then the vector field $n^a\, \=\, g^{ab} n_b$ is null on $\Delta$, where it satisfies $g_{ab} \ell^a n^b\,\= -1$. While referring to the Newman-Penrose framework, we will assume that $n_a\, \= -D_a v$ where $v$ is an affine parameter of $\ell^a$, so that $\Delta$ is foliated by $v ={\rm const}$ 2-spheres with $\ell^a$ and $n^a$ as its null normals. We can introduce a Newman-Penrose tetrad by supplementing the pair $(\ell^a, n^a)$ with a complex null vector $m^a$, tangential to these 2-spheres, satisfying $\Lie_\ell m^a\, \=\,0$ and $\eps_{ab}=2im_{[a}\bar m_{b]}$. Then the quadruplet $(\ell^a, n^a, m^a, \bm^a)$ provides the desired null tetrad on $\Delta$. Constraints on the Weyl tensor can now be made explicit as follows. Let us first pull back to $\Delta$ the indices\, $ab$ \,of $2 \nabla_{[a} \nabla_{b]} \ell^d\, \=\, - {R}_{abc}{}^d\,\ell^c$ and use the fact that $R_a{}^b \ell^a$ is proportional to $\ell^b$ on $\Delta$ to obtain:
\be \label{domega} 
\big(D_a \omega_b - D_b \omega_a\big)\ell^c \,\=  -2\, C_{\pb{ab}\,c}{}^d \ell^c \, .
\ee
Transvecting both sides with $m^d$ and $\bm^d$, one finds:
\be \label{weyl} \Psi_0 := C_{abcd}\ell^a m^b \ell^c m^d\, \=\, 0\qquad {\rm and} \qquad \Psi_1 := C_{abcd}\ell^a m^b \ell^c n^d \,\= \,0\ee
These conditions are insensitive to the specific choice of $(n^a, m^a, \bm^a)$ used to complete the null tetrad. Eq. (\ref{weyl}) also implies that $\Psi_2$ is insensitive to this choice as well. Finally, similar direct calculation shows (see, e.g., \cite{akkl1}):
\be \label{psi2} {\rm Re}\,\Psi_2\, \= -\f{1}{4} \mathcal{R} + \f{1}{24} R \qquad {\rm and} \qquad  {\rm Im}\,\Psi_2\, \epsilon_{ab}\, \=  D_{[a} \omega_{b]}\ee 
where $R$ is the 4-d scalar curvature, and $\mathcal{R}$ and $\epsilon_{ab}$ denote the pull-back to $\Delta$ of the scalar curvature $\underline{\mathcal{R}}$ and the area 2-form $\underline{\epsilon}_{ab}$ on the 2-sphere $\underline\Delta$ of null generators of $\Delta$. If there are Maxwell or Yang-Mills fields on $\Delta$, since the stress energy tensor is trace-free, $R \= 4 \Lambda$ is a constant. Therefore ${\rm Re}\Psi_2$ has essentially the same information as $\mathcal{R}$ --that determines the `shape' of the horizon-- and is time independent; $\Lie_\ell {\rm Re} \Psi_2 =0$. The shape multipoles of $\Delta$ are constructed from ${\rm Re}\Psi_2$.  Since $\Lie_\ell \omega_a =0$ it follows that ${\rm Im} \Psi_2$ is also time independent. It determines the rotational multipoles. Since $\omega_a$ is a potential for ${\rm Im} \Psi_2$, it is referred to the `rotational 1-form'. Note that the barred analogs of Eqs. (\ref{weyl}) and (\ref{psi2}) hold on all WIHs discussed in section~\ref{s2.1}. However, \emph{physical} interpretations associated with the scalar curvature $\mathcal{R}$, the 1-form $\omega_a$, and various parts of the Weyl tensor use Einstein's equations, and are thus tied to the black hole and cosmological WIHs considered in this sub-section. For example, as we noted at the end of section~\ref{s2.1}, if $\Delta$ is a WIH in a given physical space-time $(M, g_{ab})$, then it is also a WIH with respect to $g^\prime_{ab} = \mu^2 g_{ab}$ if $\Lie_\ell \mu \=0$. But if $g_{ab}$ satisfies Einstein's vacuum equations, $g^\prime_{ab}$  would not, and the $g^\prime$-multipoles of $\Delta$ will not carry any physical significance in general.

Finally, if Einstein's vacuum equations hold at $\Delta$, Eq.~(\ref{ddot}) governing the time dependence of $D$ also simplifies, providing us with a simple statement of \emph{freely specifiable} data that determines the WIH geometry. (A characterization of the free data was first obtained in \cite{abl1} using a foliation of $\Delta$). Let us first consider the case when the matter stress-energy tensor $T_{ab}$ vanishes on $\Delta$. Then Eq. (\ref{ddot}) becomes:
\ba \label{ddot2} \dot{D}_a n_b\, &\!\!:\=& \,(\Lie_\ell D_a - D_a \Lie_\ell) n_b = D_a\omega_b + \omega_a \omega_b + \ell^c\, C_{c{\,\pb{ab}}}{}^d\, n_d + \f{1}{3}\Lambda\, q_{ab}\nonumber\\
&\=& D_a\omega_b + \omega_a \omega_b \,+ {\rm Re} \Psi_2\, q_{ab} - {\rm Im} \Psi_2\, \epsilon_{ab}\, + \f{1}{3}\Lambda\, q_{ab}
\ea
We already know that $\Psi_2$ is time independent. Since $\Lie_{\ell} \omega_a \= 0$ and $\ell^a \omega_a \=0 $, it follows that $D_a \omega_b$ is the pull-back to $\Delta$ of $\underbar{D}_a \underline{\omega}_b$ on the 2-sphere $\underline{\Delta}$ of generators of $\Delta$, whence $\Lie_\ell (D_a\omega_b) \=0$. Therefore, each term on the right side is time independent and we can trivially integrate (\ref{ddot2}) to obtain:
\be \label{D} D_a n_b\, \= \,c_{ab} + \Big(D_a\omega_b + \omega_a \omega_b \,+ {\rm Re} \Psi_2\, q_{ab} - {\rm Im} \Psi_2 \,\epsilon_{ab}\, + \f{1}{3}\Lambda\, q_{ab}\Big) v \ee
where $c_{ab}$ is the integration constant;\, $\Lie_\ell (c_{ab}) \=0$. Thus, while $D$ \emph{is} time-dependent, that dependence is completely determined by the specification of $(D_a n_b)|_{v=v_\circ}$ on a 2-d cross-section $v=v_\circ$ of $\Delta$. Consequently, the free data determining the geometry $(q_{ab}, D)$ of $\Delta$ consists of the pair\,  $(q_{ab},\,  D_a n_b)|_{v=v_\circ}$ (This includes $\omega_a \,\= -\, \ell^b D_a n_b$.) In this precise sense, the local degrees of freedom in the geometry of $\Delta$ are only 2-dimensional; \emph{there are no 3-d degrees of freedom in} $D$.
\footnote{The notion of 3-d versus 2-d DOF can be traced back to the characteristic initial value problem on two intersecting 3-d null surfaces first analyzed by Sachs and Randall \cite{Sachs:1962zzb,Rendall221}. The data that can be freely specified on the 3-d null surfaces is `radiative', while the data that has to be specified on the 2-d intersection is `Coulombic'. For any given radiative data, the value of the Coulombic fields along the null hypersurfaces is uniquely determined by Einstein's equations. Hence they correspond to 2-d degrees of freedom
or `corner data'. On $\Delta$ there are no 3-d degrees of freedom at all; the radiative fields vanish and the 2-d fields are transported to full $\Delta$ by (\ref{ddot2}).} 
Note that the lack of 3-d degrees of freedom is a consequence of Einstein's equations; it embodies the intuition that there is no flux of gravitational radiation across $\Delta$. We will see in section \ref{s2.3} that the situation is very different if \emph{conformal} Einstein's equations are satisfied at the WIH.  

Let us now allow for Maxwell fields as sources. Then condition (iii) of Definition 1 implies that the stress-energy tensor $T_{ab}$ is such that $T_{ab} \ell^b \propto \ell_b$ on $\Delta$. Since $T_{ab}\ell^a\ell^b \=0$, it follows that the `radiative part'\, $\Phi_0 \= F_{ab} \ell^a m^b$\, of the Maxwell field on $\Delta$ vanishes identically. Maxwell's equations then imply that the other 4 components $\Phi_1$ and $\Phi_2$ are determined by their values on a cross-section $v\=v_\circ$ of $\Delta$. Thus there are no 3-d degrees of freedom in the Maxwell field either. Returning to the gravitational free data on $\Delta$, in presence of a Maxwell field Eq. (\ref{ddot2}) acquires an extra term on the right side but it is again time independent. Hence the free data determining geometry of $\Delta$ continues to consist of 2-d fields, specified on a cross-section. Situation with Yang-Mills fields is completely analogous.

To summarize, although the intrinsic derivative operator $D$ is time \emph{dependent} on generic black hole and cosmological WIHs $\Delta$, there are no 3-d radiative degrees of freedom either of the gravitational field nor of Maxwell or Yang-Mills fields. This is why these WIHs have been regarded as representing surfaces whose geometry is `in equilibrium'. Note, however, that this notion of equilibrium is much weaker than that embodied in the notion of Killing horizons. Indeed, as mentioned in section \ref{s3.1} one can always endow an NEH with a WIH structure, but \emph{not} an IH structure, and Killing horizons are even more restrictive than IHs. In particular, there exist explicit examples --the Robinson-Trautmann vacuum solutions and Kastor-Traschen electro-vac solutions \cite{pc,Kastor1993}-- that admit an IH that is not a Killing horizon. Furthermore, using the characteristic initial value problem, Lewandowski \cite{Lewandowski:1999zs} has shown that a generic IH is not a Killing horizon. These results show that a widespread belief that WIHs are `only slightly weaker than Killing horizons' is misplaced. 

Finally, the weak notion of equilibrium encapsulated by $\Delta$ was directly traced back to the fact that 
$g_{ab}$ satisfies Einstein's equations with Maxwell or Yang-Mills sources on $\Delta$.\, $\scrip$, on the other hand, is a WIH in the conformally completed space-time $(\hM,\hg_{ab})$ and $\hg_{ab}$ does not satisfy this condition. Therefore, as we now show, its geometry does not carry \emph{any} connotation of being in `equilibrium'. Even though it \emph{is} a WIH, the physics of $\scrip$ is very different from that on $\Delta$ considered in this subsection.

\subsection{$\scrip$ as a WIH}
\label{s2.3}

Let us then consider the complementary case where $(\bM, \bg_{ab})$ of section \ref{s2.1} is taken to be the conformal completion $(\h{M}, \h{g}_{ab})$ of an asymptotically flat space-time and $\WIH$ is replaced by $\scrip$. In this section we will set $\Lambda=0$ since $\scrip$ is null only in this case. We will show that $\scrip$ is then a WIH, and explain why its physics is nonetheless very different from that of black hole and cosmological horizons $\Delta$. 

Let us begin by recalling from \cite{doi:10.1063/1.524467} the notion of asymptotic flatness  that is relevant to our analysis.\\
{Definition 2:} A physical space-time $(M, g_{ab})$ will be said to be asymptotically flat (at null infinity) if there exists a Manifold $\hM$ with boundary $\scrip$ equipped with a metric $\hg_{ab}$ such that $\hM = M\cup \scrip$ and $\h{g}_{ab} = \Omega^2 g_{ab}$ on $M$ for a nowhere vanishing function $\Omega$, and,\\
(i) $\scrip$ has topology $\mathbb{S}^2\times \mathbb{R}$ and its causal past contains a non-empty portion of $M$;\\
(ii) $\Omega \= 0$, where $\=$ will now refer to equality at points of $\scrip$\!,\, while \,$\h\nabla_a \Omega$\, is nowhere vanishing on $\scrip$; and,\\
(iii) The physical metric $g_{ab}$ satisfies Einstein's equations $R_{ab} - \f{1}{2} R g_{ab} = 8\pi G\, T_{ab}$ where $\Omega^{-1} T_{ab}$ admits a smooth limit to $\scrip$.\\
$(\hM, \hg_{ab})$ will be referred to as the Penrose completion of the physical space-time $(M, g_{ab})$.\\
The asymptotic condition (iii) on the stress energy tensor is the standard one; it is satisfied by matter sources normally used in general relativity, in particular, the Maxwell field. 

Let us set $\hn_a := \h\nabla_a \Omega$. Then, by rewriting the field equations in terms of the conformally rescaled metric, we obtain:
\be \label{fe} \hR_{ab} - \textstyle{\f{1}{2}} \h{R}\, \h{g}_{ab} + 2\,\Omega^{-1} \big(\h\nabla_a \hn_b - (\h\nabla^c\hn_c)\, \h{g}_{ab}\big) + 3\,\Omega^{-2}\, (\hn^c \hn_c)\,\hg_{ab} = 8\pi G T_{ab}
\ee
As is well-known, Eq. (\ref{fe}) and the assumption (iii) on the fall-off of stress-energy tensor have several important implications (see, e.g., \cite{gerochrev,aa-yau}):

(1) $\hn^a\, \hn_a \=0$.\, Since $\hn_a$ is normal to $\scrip$, this implies that $\scrip$ is a null 3-manifold.  \smallskip

(2) There is considerable conformal freedom in the Penrose completion. Given a completion satisfying condition (iii) of Definition 2, $\Omega^\prime = \mu \Omega$ is also an allowable conformal factor, if $\mu$ is nowhere vanishing on $\scrip$. One can use this freedom to set $\h\nabla_a\hn^a \= 0$. Then the restricted conformal freedom is $\Omega \to \mu \Omega$, with $\Lie_{\hn} \mu \=0$. This choice of divergence-free conformal frame can be made without any loss of generality and removes unnecessary complications in calculations that can obscure the geometrical and physical significance of various fields and properties. Therefore it is widely used in the literature on null infinity.%

\emph{We will restrict ourselves to these divergence-free conformal frames.} With this restriction, (\ref{fe}) further implies that the full derivative of $\hn^a$ vanishes at $\scrip$: \, $\h\nabla_a \h{n}^b \= 0$.\smallskip 

(3) The Schouten tensor $\hS_a{}^b = \hR_a{}^b - \f{1}{6}\hR\, \delta_a{}^b$ of $\hg_{ab}$ satisfies: $\hS_a{}^b \hn^a \= -\hf \hn^b$, with $\h{f} = \Omega^{-2}\hn^a \hn_a$. Therefore condition (ii) in {Definition 2} is satisfied: $\hR_{a}{}^b \hn^a \= \alpha\, \hn^b$ with $\alpha \= \f{1}{6}\hR -\hf$.\smallskip

(4) The Weyl tensor $\hC_{abc}{}^d$ of $(\hM, \hg_{ab})$ vanishes on $\scrip$. Hence if $\h{g}_{ab}$ is $C^3$ at $\scrip$, then $\hK_{abcd} := \Omega^{-1} \hC_{abcd}$ admits a continuous limit to $\scrip$.
\vskip0.2cm
Let us examine the consequences of these properties of fields at $\scrip$. Property (1) ensures that $\scrip$ is a \emph{null} 3-dimensional manifold in the conformally completed space-time $(\hM, \hg_{ab})$ with $\hn^a$ as a normal; (2) implies that expansion $\theta_{\hn}$ of $\hn^a$ vanishes on $\scrip$; and (3) implies that $\hR_a{}^b$ satisfies condition (iii) in Definition 1. Therefore, $\scrip$ is an NEH in $(\hM, \hg_{ab})$ in every divergence-free conformal completion of the given physical space-time.\smallskip

Let us denote by $\hD$ the derivative operator induced by $\h\nabla$ on $\scrip$. From our discussion of section \ref{s2.1} it follows that $\hD_a \hn^b \= \h\omega_a \hn^b$. However, we know from (2) that $\h\nabla_a \hn^b \= 0$. Hence, the 1-form $\h\omega_a$ on $\scrip$ in fact \emph{vanishes} for all asymptotically flat space-time. In particular, $\h\omega_a$ is divergence-free. Thus $\h{n}^a$ is a preferred geodesic null normal on the NEH $\scrip$, and since $\Lie_{\hn} \omega_a \=0$ trivially,\, it follows that $\scrip$ endowed with $\hn^a$ is a WIH. Note that on a general WIH, we only have an equivalence class $[{\bar{k}}^a]$ of preferred null normals with this property, where two are equivalent if they differ via rescaling by a positive constant. At $\scrip$, the given conformal completion naturally selects for us canonical vector field in this class: $\hn^a := \hg^{ab} \h\nabla_a \Omega$. If we use $\Omega^\prime = c\, \Omega$ for a positive constant $c$, we obtain another space-time $(\hM, \hg^\prime_{ab} = c^2 \hg_{ab})$. Even though the two completions refer to the same \emph{physical} space-time, in {Definition 1} they have to be regarded as distinct space-times that share a WIH.

More generally, $(\hM, \hg_{ab})$ and $(\hM, \hg^\prime_{ab} = \mu^2 \hg_{ab})$, where $\mu$  (is nowhere vanishing on $\scrip$ and)  satisfies $\Lie_{\hn} \mu \=0$ present us with two divergence-free conformal completions of the same physical space-times. But in the WIH perspective, they are to be regarded as distinct space-times, in both of which $\scrip$ is a WIH but with distinct NEH geometries $(\hq_{ab},\hn^a, \hD)$ and $(\hq_{ab}^\prime,\, \hn^{\prime\,a}, \hD^\prime)$. Put differently, while the fact that $\scrip$ is an extremal WIH is a property of the given physical space-time, the geometry of this WIH varies from one (divergence-free) conformal completion to another. Now, in the discussion of asymptotic symmetries and associated charges and fluxes, one has to consider not just a fixed $\hg_{ab}$ but the entire conformal class of space-times and ensure that physical results are insensitive to this conformal freedom. We will do so, mirroring the procedure followed in the standard treatment of $\scrip$ where one first works with a fixed conformal completion and then shows that the results are insensitive to this choice.

Let us return to Eq. (\ref{ddot}) satisfied on \emph{general} WIHs\, $\WIH$\, of section~\ref{s2.1}:
\be \label{ddot3} \dot{\bD}_a \j_b \= \bD_a\bomega_b + \bomega_a \bomega_b + {\bar{k}}^c\, \bC_{c\,{\pb{ab}}}{}^d\, \j_d + \textstyle{\f{1}{2}}\,\big(\bS_{\pb{ab}} + (\alpha-\frac{1}6\bar R) \bq_{ab}\big)\, . \ee
As we saw in section~\ref{s2.2}, on black hole and cosmological horizons $\Delta$, the Ricci part (encoded in \,$S_{ab}$\, and\, $\alpha$) is non-dynamical and in fact vanishes if vacuum equations hold (with $\Lambda=0$). Therefore dynamics of $D$ is governed by $\omega_a$ and $C_{c\,{\pb{ab}}}{}^d$. At $\scrip$, the situation is just the opposite! Now both $\h\omega_a$ and $\hC_{abcd}$ vanish while the Ricci part $\h{R}_{ab}$ of $\hat{g}_{ab}$ is now non-zero even when the physical metric satisfies vacuum Einstein's equations (with $\Lambda=0$). Therefore dynamics of $\hD$ is governed by the Ricci part, precisely the term that vanishes on $\Delta$ when vacuum Einstein's equations hold.

For agreement with the terminology used in the literature on $\scrip$, let us replace $\j_b$ by a 1-form $\h\ell_a$ on\, $\scrip$\, satisfying $\hn^a\h\ell_a \,\=-1$. Then we have:
\be \label{ddot4} \dot{\hD}_a \h\ell_b\, \=\, \f{1}{2}\,\big(\hS_{\pb{ab}} + (\alpha-\frac16\hat R)\hq_{ab}\big) \ee
where $\alpha \= \f{1}{6}\hR - f \equiv \f{1}{6}\hR - \lim_{\scrip} \Omega^{-2} \hn^a\hn_a$. Now, as is well-known (see, e.g., \cite{gerochrev,aa-radiativemodes,aams,aa-yau}), the Bondi news resides in $\hS_{\pb{ab}}$: It is given by $\h{N}_{ab} \= {\hS}_{\pb{ab}} -\h\rho_{ab}$ where $\h\rho_{ab}$ is the kinematical Geroch tensor field \cite{gerochrev} that serves to remove the unphysical information contained in $\hS_{\pb{ab}}$ through our choice of $\Omega$. Therefore, not only is $\hD$ time-dependent at $\scrip$, but the time-dependence is now encoded in Bondi-news -- \emph{a field with two degrees of freedom per point in 3 dimensions!} Thus, in sharp contrast to what we found in section~\ref{s2.2} for $\Delta$, the freely specifiable data in the WIH geometry\, $(\h{q}_{ab},\hn^a, \h{D})$\, of $\scrip$ includes a field with \emph{3-d degrees of freedom}. While the connection $D$ on $\Delta$ can be specified by providing fields just on a cross-section of $\Delta$, this is not possible at $\scrip$ because the curvature of $\hD$ contains the Bondi news $\h{N}_{ab}$ which can be freely specified on entire $\scrip$. 

To summarize, $\Delta$ and $\scrip$ are WIHs and therefore they both inherit the rich structure discussed in section \ref{s2.1}, as well as the dynamical equation (\ref{ddot3}). Still they contain \emph{very} different physics because complementary terms of Eq. (\ref{ddot3}) trivialize in the two cases. In vacuum space-times (with $\Lambda=0$) the Ricci contribution vanishes and the time dependence of $D$ is dictated by $\bomega_a$ and $\bC_{c\,{\pb{ab}}}{}^d$ on $\Delta$ (both of which are themselves time independent). By contrast, both these terms vanish identically at $\scrip$ and the time dependence of $\hD$ is now dictated by $\hR_{\pb{ab}}$ (which has 3-d degrees of freedom).  Thus every term that contributes to the time derivative of $D$ on $\Delta$ vanishes on $\scrip$ and vice versa! It is because of this subtle and surprising `complementarity' that both $\scrip$ and $\Delta$ can be realized as distinct special cases of WIHs, sharing in detail a large number of properties, and yet carrying very different physics.
\section{Universal Structure and Symmetry Groups}
\label{s3}
\subsection{General WIHs\, $\WIH$}
\label{s3.1}

Let us begin by recalling (from \cite{akkl1}) the universal structure and the symmetry group of a general  WIH \,$\WIH$. Each  WIH is equipped with an intrinsic metric $\bq_{ab}$ and an equivalence class of preferred null normals $[{\bar{k}}^a]$. However, these fields are not universal; for example, the curvature of $\bq_{ab}$ varies from one black hole  WIH to another. On the other hand, each  WIH admits a 3-parameter family of \emph{unit, round} 2-sphere metrics $\qo_{ab}$ that are conformally related to its $\bq_{ab}$: $\qo_{ab} \= \psio^2 \bq_{ab}$.%
\footnote{More precisely, the 2-sphere $\ul\WIH$ (of null generators of $\WIH$) admits a unique 3-parameter family of unit round metrics $\qoub_{ab}$ conformally related to $\ub{q}_{ab}$, and $\qo_{ab}$ are the pull-backs to $\WIH$ of $\qoub_{ab}$. For brevity, we will not explicitly mention $\ul\WIH$ and freely pass between fields on $\ul\WIH$ and their pull-backs to $\WIH$.}
While the conformal factors $\psio$ relating the metrics $\qo_{ab}$ and $\bq_{ab}$ vary from one  WIH to another, the \emph{relative} conformal factors $\alphao$ between any two round metrics \emph{is universal}:
\be \label{trans} \qo^\prime_{ab} \= \alphao^2 \qo_{ab}\quad {\hbox{\rm where $\alphao$ satisfies}} \quad
\Do^2 \ln \alphao + 1\, \=\, \alphao^{-2}.\ee
$\alphao$ is time-independent since all 2-sphere metrics are, and the last equation admits precisely a 3-parameter family of solutions, given by 
\ba \label{alpha}
\alphao^{-1} &=& \alpha_0 + \sum_{i=1}^3 \, \alpha_i\, \hat{r}^i\, , \quad{\hbox{\rm where $\alpha_0$ and $\alpha_i$ are real \emph{constants},\,\, satisfying}} \nonumber\\
&-& \alpha_0^2 + \sum_{i=1}^3 (\alpha_i)^2 = -1 , \qquad {\rm and}\qquad \hat{r}^i = (\sin\vartheta\cos\varphi,\, \sin\vartheta\sin\varphi,\, \cos\vartheta)\, . \ea   
Here $(\vartheta, \varphi)$ are spherical polar coordinates adapted to the first round metric $\qo_{ab}$. Motivated by the transformation properties of vectors under conformal rescalings, and also by multipole moment considerations,%
\footnote{With this rescaling, the 1-form $\omegao_a$ defined by $D_a\ko^b = \omegao_a\ko^b$ can be written as $ \omegao_a = -\big(\Do_a E + \epsilono_a{}^b \Do_b B\big)$, where $E$ and $B$ serve as scalar potentials for the real and imaginary part of $\Psi_2$. The shape and rotational multipoles are obtained by a spherical harmonic decomposition of $E$ and $B$ defined by $\qo_{ab}$.} 
one carries out a parallel rescaling of the vector fields $[{\bar{k}}^a]$, and sets $[\ko^a] \= [\psi^{-1}{\bar{k}}^a]$\,\,\cite{akkl1}. Then \emph{every}  WIH is naturally equipped with a 3-parameter family of pairs $(\qo_{ab},\, [\ko^a])$ related by conformal factors $\alphao$ of Eq. (\ref{alpha}): $(\qo^\prime_{ab},\, [\ko^{\prime\,a}])\, \=\, (\alphao^2\, \qo_{ab},\, \alphao^{-1}\, [\ko^a])$. The relation is the same for all  WIHs. Thus, the pairs capture the `kinematical' structure that is universal, leaving out fields such as $\bq_{ab}, [{\bar{k}}^a], \bomega_a$ carrying (physical) information that varies from one  WIH to another. 

In view of this universal structure shared by all  WIHs, we are led to introduce an abstract 3-manifold $\WIHo$, topologically $\mathbb{S}^2\times\mathbb{R}$, equipped with a 3-parameter family of pairs $(\qo_{ab}, [\ko^a])$ of unit round metrics $\qo_{ab}$, and equivalence classes of vector fields $\ko^a$, where two are equivalent if they are rescalings of one another with a positive constant $c$, such that:\smallskip

\noindent(i) the space of integral curves of the chosen vector fields $\ko^a$ is diffeomorphic to $\mathbb{S}^2$, and any two vector fields are related by $\ko^{\prime\,a} \= c\, \alphao^{-1}\, \ko^a$, where $c$ is a positive constant and functions $\alphao^{-1}$ are given by (\ref{alpha});\, and,\\
(ii) any two metrics are related by $\qo^\prime_{ab} \= \alphao^2 \qo_{ab}$.\smallskip

\noindent Thus, the 3-manifold $\WIHo$ is equipped with the structure that is common to all  WIHs. The symmetry group\, $\G$ \,of\,  WIHs\, is the subgroup of the diffeomorphism group of \,$\WIHo\,$ that preserves this universal structure. Given any concrete  WIH\, $\WIH$\,  that resides in a space-time $(\bM, \bg_{ab})$, there is a diffeomorphism  from \,$\WIH$\, to $\WIHo$ that sends the pairs $(\qo_{ab}, [\ko^a])$ on \,$\WIH$\, to those we fixed on $\WIHo$. Any two such diffeomorphisms are related by an  WIH symmetry in $\G$. \vskip0.1cm

By definition, $\G$ is generated by vector fields $\xi^a$ on $\WIHo$, whose action preserves the universal structure: Their defining property is:
\be \label{symmetry} \Lie_\xi\, \qo_{ab} \= 2\betao\, \qo_{ab} \qquad {\rm and} \qquad \Lie_\xi {\ko}^a \= - (\betao +\varpio ){\ko}^a\ee
where $\varpio$ is a constant and $\betao$ satisfies $\big(\Do^2 +2\big)\, \betao =0$; it is a linear combination of the first three spherical harmonics of $\qo_{ab}$. $\G$ has a rich structure that has been spelled out in \cite{akkl1}, and is a subgroup of the symmetry group of a general null hypersurface studied in \cite{cfp} (see also \cite{Donnay:2015abr,Hopfmuller:2018fni,Chandrasekaran:2021hxc,Freidel:2021yqe,Adami:2021nnf,Adami:2021kvx,Odak:2023pga,Chandrasekaran:2023vzb,Ciambelli:2023mir,Adami:2023wbe}). 
We will focus just on those aspects that are needed to understand the relation between symmetries of general  WIHs\, $\WIH$\, and those of\, $\scrip$. A key property of $\G$ is that, as we show below, it admits a \emph{normal subgroup} $\B$ that is isomorphic to the BMS group, and the quotient $\G/\B$ is a 1-d Lie group $\mathfrak{D}$ of `dilations'. Thus $\G = \B \ltimes \mathfrak{D}$; \,it has one more generator than the BMS group. This is because, as explained in section \ref{s3.2}, the universal structure on a general WIH $\WIH$ is slightly weaker than that on $\scrip$.

To explore the structure of $\G$, let us begin with an explicit expression of generators $\xi^a$ defined by (\ref{symmetry}). Fix a round metric $\qo_{ab}$, the associated $[\ko^a]$ in the universal structure on $\WIHo$, and a cross-section $C$ of $\WIHo$. Let $\vo$ denote the affine parameter of an $\ko^a \in [\ko^a]$ with $\vo=0$ on $C$. Finally, introduce a set of spherical coordinates $\vartheta,\varphi$ defined by $\qo_{ab}$ on $C$ and Lie-drag them by $\ko^a$. Thus, $\WIHo$ is now endowed with a global chart $(\vo,\vartheta, \varphi)$. In terms of this structure, the general symmetry vector field $\xi^a$ can be expanded out as:
\be \label{xi1} \xi^a \= \big((\varpio + \betao)\vo + \mathring{\mathfrak{s}} \big) \ko^a \,+\, \epsilono^{ab} \Do_b \chio\,  - \, \qo^{ab} \Do_b \betao\, , \ee
where $\mathring{\mathfrak{s}}(\vartheta,\varphi)$ is a general function on the 2-sphere of null generators of $\WIHo$; $\chio(\theta,\varphi)$ and $\betao(\vartheta,\varphi)$ are both linear combinations of first three spherical harmonics defined by $\qo_{ab}$; and  $\epsilono^{ab},\, \qo^{ab}$ are the inverses of the area 2-form and the metric on the $\vo = {\rm const}$ cross-sections, respectively. Given this explicit form, one can readily check that $\xi^a$ satisfies Eqs.~(\ref{symmetry}) that define infinitesimal WIH symmetries. Conversely, given a pair $(\qo_{ab}, \ko^a)$ from the universal structure, every symmetry vector $\xi^a$ has this form.
(Note, incidentally, that a sign error has percolated in several papers on the subject, including \cite{Ashtekar:2024mme}, where the last term in (\ref{xi1}) appears with a positive sign.)

The terms in the decomposition can be interpreted as follows:\, $d^a := \varpio\, \vo\, \ko^a$ is a dilation;\, $S^a:= \mathring{\mathfrak{s}} \ko^a$ a supertranslation;\, $R^a := \epsilono^{ab} \Do_b \chio$ a rotation; and $B^a := \qo^{ab} \Do_b \betao\, - \vo \betao\, \ko^a$\, a boost. Thus, $\xi^a\, =\, (d^a + S^a) + R^a + B^a\, $. As in the BMS case the descriptors $\mathring{\mathfrak{s}}(\vartheta,\varphi)$ of supertranslations $S^a$ are conformally weighted functions: under $(\qo_{ab},\, [\ko^{a}]) \,\to\, (\alphao^2\, \qo_{ab},\, \alphao^{-1}\, [\ko^a])$\, we have $\mathring{\mathfrak{s}} \,\to\, \alphao\, \mathring{\mathfrak{s}}$; the rest is invariant. The Lie algebra $\g$ of horizon symmetries admits a 4-dimensional \emph{Lie-ideal} of `translations' consisting of those supertranslations whose descriptors $\mathring{\mathfrak{s}}(\vartheta,\varphi)$ are linear combinations of the first four spherical harmonics of any round metric $\qo_{ab}$ on\, $\WIHo$\, in our collection. Under $\qo_{ab} \to \alphao^2 \qo_{ab}$, this 4-dimensional space is left invariant, just as in the BMS Lie-algebra $\LBMS$.  While these features are shared by $\g$ and $\LBMS$, the  WIH symmetry Lie algebra $\g$ contains an extra generator: the dilation\, $\xi^a = \varpio\,\vo\,\ko^a$\, that is a vertical vector field like supertranslations, \emph{but depends on $\vo$ rather than on} $(\vartheta,\,\varphi)$. 

The decomposition (\ref{xi1}) of the general symmetry vector field is tied to the choices we made to provide its  explicit expression. If we keep $(\qo_{ab}, [\ko^a])$ the same but change the cross-section $C$, then the affine parameter $v_o$ will change via $v_o \to v_o + f(\vartheta,\varphi)$ and the new dilation vector field would be a linear combination of the old dilation and a super-translation.%
\footnote{On the other hand, if we keep the cross-section the same but change $\ko^a$ via $\ko^a \to \alphao^{-1}\ko^a$,  then the dilation $d^a$ is left unchanged  because $\vo \to \,\alphao v_o$.} 
Thus, while the notion of a pure supertranslation is well-defined --i.e. does not depend on the choice of the auxiliary structure we introduced-- the notion of a `pure' dilation is not. Put differently, while supertranslation vector fields constitute a canonical sub-algebra of $\g$, dilations are in 1-1 correspondence with the elements of the quotient $\g/\LBMS$ of the Lie algebra $\g$ by its BMS sub-algebra $\LBMS$; one needs extra structure to embed them in the Lie algebra of $\G$. These properties of the Lie algebra of $\G$ imply that, as noted above, $\G$ admits $\B$ as a normal subgroup, with $\G/\B = \mathfrak{D}$, the 1-dimensional group of dilations. Finally, as in the BMS group, the choice of the cross-section $C$ provides us with a preferred Lorentz subgroup of $\G$ generated by the six symmetry vector fields $\xi^a$ that are tangential to $C$. Therefore, under $\vo \to \vo + f(\vartheta,\varphi)$ this subgroup changes by a super-translation, just as in the BMS group. 

Finally note that one does not acquire any extra structure in the passage from general  WIHs $\WIH$ to the black hole and cosmological horizons $\Delta$. Therefore the symmetry group and its detailed structure discussed above \emph{is the same for $\Delta$ as it is for general  WIHs $\WIH$.} However, as we show in section \ref{s3.2}, the situation at $\scrip$ is different.\\

\emph{Remark:} There is a conceptual point about the notion of symmetries that one has to keep in mind. Symmetries of a given space-time are generated by a Killing vector and a generic metric $\bg_{ab}$ does not admit any Killing vectors. Nonetheless, the space of asymptotically flat space-times does admit a symmetry group; it preserves only the asymptotic universal structure that is shared by the collection of space-times satisfying the specified asymptotic conditions. Similarly, a generic metric $\bg_{ab}$ admitting  WIH $\WIH$ will not carry any vector field $\xi^a$ whose action preserves $\bg_{ab}$ even at $\WIH$. What the symmetry  generators $\xi^a$ discussed above preserve is only the universal structure that is common to all space-times that admit an  WIH $\WIH$.

\subsection{$\scrip$ as a special case of\, $\WIH$}
\label{s3.2}

Each divergence-free conformal completion $(\hM, \hg_{ab})$ of an asymptotically flat space-time endows $\scrip$ with an intrinsic (degenerate) metric $\hq_{ab}$ and a null normal $\hn^a$ and as one changes the conformal factor $\Omega$, these fields on $\scrip$ change via $(\hq_{ab},\, \hn^a)\, \to  \, (\mu^2 \hq_{ab},\, \mu^{-1} \hn^a)$, where $\mu$ satisfies $\Lie_{\hn} \mu =0$. Universal structure of $\scrip$ is generally taken to consist these pairs $(\hq_{ab}, \hn^a)$ (see, e.g., \cite{schmidt,gerochrev,aa-yau,ashtekar1987asymptotic}). From the geometrical  WIH perspective of section \ref{s2.1}, on the other hand, the intrinsic metrics $\bq_{ab}$ on \, $\WIH$\, are \emph{not} a part of the universal structure because, e.g., they vary from one black hole or cosmological horizon to another, encoding physical properties that distinguish these horizons from one another. Similarly, the symmetry group $\G$ is not isomorphic to the BMS group $\B$ but to a 1-d extension thereof. Thus, at first, there seems to be a tension between the two perspectives. We will now show that this tension is only apparent: If one restricts oneself to the \emph{sub-class} of  WIHs that are boundaries $\scrip$ in Penrose conformal completions of asymptotically flat space-times, then the  WIHs acquire a small additional structure that reduces $\G$ to the BMS group $\B$, \emph{with the familiar concrete action} on $\scrip$.

Let us begin by recalling a key feature of general  WIHs. As discussed in section~\ref{s2.1}, in general we can single out only an \emph{equivalence class} $[{\bar{k}}^a]$ of preferred geodesic null normals to \,$\WIH$ --where two differ from each other by rescaling with a positive constant-- rather than a specific null normal ${\bar{k}}^a$. It is this fact that forces one to allow dilations --which are absent in the BMS group-- as  WIH symmetries. Could this be because we just left out some structure on $\WIH$ inadvertently? The answer is in the negative. Additional structure to select a preferred element of this equivalence class simply does not exist on a general  WIH \cite{akkl1}. For example, the action of the static Killing field $t^a$ in the Schwarzschild space-time rescales our geodesic null normals ${\bar{k}}^a$ on the Schwarzschild horizon by a positive constant. Since the action of the Killing field leaves the entire 4-geometry invariant, already in this case one cannot invariantly single out a preferred ${\bar{k}}^a$. 

The situation changes when we restrict the general $(\bM,\, \bg_{ab})$ to be conformal completions $(\hM,\,\hg_{ab})$ of physical space-times $(M,g_{ab})$, and focus on the  WIHs provided by $\scrip$. Since each completion comes with a conformal factor $\Omega$, it provides a \emph{canonical} null normal $\h{n}^a\, \=\, \hg^{ab}\nabla_a \Omega$ to the  WIH\, $\scrip$, rather than an equivalence class $[\h{n}^a]$. Recall that on a general  WIH, the unit round metrics $\qo_{ab}$ are conformally related to the $\bq_{ab}$ induced by physical metrics $\bg_{ab}$ via $\qo_{ab}\, \=\, \psio^2 \bq_{ab}$, and come paired with null normals $\ko^a$ that are related to the null normals $[{\bar{k}}^a]$ selected by $\bar{g}^{ab}$ via $[\ko^a]\, \=\, [\psio^{-1} {\bar{k}}^a]$. Since there is now a canonical $\hn^a$, we have a 3-parameter family of pairs $(\hqo_{ab}, \hno^a)$ that feature vector fields $\hno^a$, rather than equivalence classes $[\hno^a]$. Any two pairs are again related by $(\hqo{}^\prime_{ab}, \hno^{\prime\, a})\, \= \,(\alphao^2\hqo_{ab}, \alpha^{-1}\hno^a)$, with $\alphao$ given by (\ref{alpha}). The  WIH perspective puts these pairs on the forefront. 
By contrast, the emphasis in the usual discussions of universal structure is on $(\hq_{ab},\, \hn^a)$, where $(\hq_{ab}^\prime,\, \hn^{\prime\,a})\, \=\, (\mu^2 \hq_{ab},\, \mu^{-1} \hn^a)$ and $\mu$ only satisfies $\Lie_{\h{n}}\, \mu \=0$) (see, e.g., \cite{schmidt,gerochrev,aa-yau}).  In spite of this shift of emphasis, as we now show, the  group of asymptotic symmetries is again the BMS group. 

For the restricted class of  WIHs provided by $\scrip$, vector fields $\xi^a$ on the abstract  WIH $\WIHo$\, are symmetries if the diffeomorphisms they generate preserve the collection of pairs $(\hqo_{ab}, \hno^a)$. Consequently, they must satisfy a slightly stronger version of (\ref{symmetry}):
\be \label{symmetry2} \Lie_\xi\, \hqo_{ab} \= 2\betao\, \hqo_{ab} \qquad {\rm and} \qquad \Lie_\xi {\hno}^a \= - \betao \,{\hno}^a\, ; \ee
the constant $\varpio$ in (\ref{symmetry}) is now set to zero because the rescaling freedom in
each $[\ko^a]$ no longer exists. Put differently, on a general  WIH, a symmetry vector field $\xi^a$ can leave each metric $\qo_{ab}$ invariant, but rescale the null normals $\ko^a$ by a constant $\varpio$ (see Eq.~(\ref{symmetry})). That is no longer possible; now rescalings of $\hqo_{ab}$ and of $\hno^a$ must occur in tandem. Thus, now symmetry generators are given by setting $\varpio =0$ in (\ref{xi1}):
\be \label{xi2} \xi^a \= \big(\mathring{\mathfrak{s}}\,+\, \betao\,\uo \big) \hn^a \,+\, \hepsilono^{ab} \Do_b \chio\, - \, \hqo^{ab} \Do_b \betao\,  \ee
where $\uo$ is now an affine parameter of the null normal $\hn^a$. By inspection, these are the standard BMS vector fields written out in a `Bondi conformal frame' $(\hqo_{ab}, \hno^a)$ \cite{PhysRevD.101.044005}, and their action on $\scrip$ regarded as an  WIH is the same as the familiar one without the  WIH perspective. 

Consequently, the symmetry group $\G$ of general  WIHs and their concrete action on \,$\WIHo$\, reduces to the BMS group $\B$ with its standard action on $\scrip$. The difference with treatments of $\scrip$ that focus on the pairs $(\hq_{ab},\, \hn^a)$ induced on $\scrip$ by $\hg_{ab}$ and $\h\nabla_a\Omega$ is that the  WIH perspective puts the Bondi conformal frames $(\hqo_{ab},\, \hno^a)$ at the forefront. It is these pairs, with $\hqo_{ab}$ the unit, round 2-sphere metrics, that specify the universal structure now. In this sense, for $\scrip$, the notion of symmetries that descends from the  WIH perspective is somewhat more closely related to the way it was first introduced by Bondi, Metzner and Sachs \cite{bondi1962mgj,sachs} (although one now works in conformally completed space-times and makes no reference to to coordinate systems). 

To conclude, let us recall  that inclusion of the dilation symmetry $\xi^a \= \varpio\,\vo\, \ko^a$ in the Lie algebra $\g$ is essential in the case of black hole and cosmological horizons because in the case when the  WIH is a non-extremal Killing horizon, the restriction of the Killing vector field to the  WIH is a dilation \cite{akkl2}. Had the dilation been absent, the framework would have implied that Killing symmetries of the full space-time $(\bM, \, \bg_{ab})$ that are tangential to \,$\WIH$\, need not be  WIH symmetries! Is there an analogous problem at $\scrip$ now, since dilation is no longer a symmetry of this  WIH? The answer is in the negative: every Killing symmetry of the physical space-time admits a well-defined limit to $\scrip$ which is tangential to $\scrip$ and belongs to the BMS Lie algebra $\LBMS$ \cite{doi:10.1063/1.524467}. In particular, if an asymptotically flat space-time is stationary and admits a non-extremal black hole  WIH, then the restriction of the Killing field to the black hole  WIH is a dilation, while its restriction to the $\scrip$  WIH belongs to the translation sub-Lie-algebra of $\LBMS$. Thus, there is an interesting and rather subtle interplay between what happens at horizons in the physical space-time and at $\scrip$, allowing one to consistently treat both in a single, general  WIH framework. 
\\

\emph{Remark:} Recall from the Remark at the end of section \ref{s3.1} that an  WIH-symmetry does not preserve the metric $\bar{g}_{ab}$ on general  WIHs $\WIH$, but only the conformal class of intrinsic metrics. What is the situation for $\scrip$? To regard $\scrip$ as an  WIH we have to fix a conformal completion $(\hM, \hg_{ab})$ of an asymptotically flat space-time $(M, g_{ab})$ and the space-time extension of a general BMS symmetry does not preserve the given $\h{g}_{ab}$ even at $\scrip$, but only the conformal class of intrinsic metrics. Thus, for the metric the situation at $\scrip$ is the same as that for  $\Delta$. The only difference is that the rescaling of the metric and the null normal are not as tightly intertwined at $\Delta$ as they are at $\scrip$.

In both cases, to arrive at charges and fluxes associated with symmetries, we have to work with a phase space which carries a well-defined action of the symmetry group --i.e., all space-times that admit the desired  WIH as a boundary. As we will see in there companion paper \cite{aass2}, if the phase space includes a $(\hM, \hg_{ab})$ that arises from a divergence-free conformal completion of a physical space-time $(M, g_{ab})$, it will include all other divergence-free completions as well.

\section{Discussion}
\label{s4}

The fact that null infinity is a WIH seems very surprising at first because WIHs are commonly associated with black hole (and cosmological) horizons $\Delta$. Indeed, $\scrip$ and $\Delta$ have almost the opposite connotations. $\scrip$ lies in the asymptotic, weak curvature region, while $\Delta$ lies in a strong field region. $\scrip$ is the arena for discussing gravitational waves, while $\Delta$ is generally used to discuss the Coulombic properties of black holes in equilibrium. The goal of this paper was to probe this tension. We began in  section \ref{s2} by showing that the  resolution lies in three facts that are rather subtle: 
\begin{itemize} 
\item First, the general notion of a weakly isolated horizon $\WIH$ refers only to a metric, say $\bg_{ab}$, of signature -,+,+,+ on a manifold $\bM$. The notion does not require $\bg_{ab}$ to satisfy either Einstein's equations or conformal Einstein's equations. The only condition on the Ricci tensor $\bR_a{}^b$ is that, if ${\bar{k}}^a$ is a null normal to $\WIH$, then $\bR_a{}^b {\bar{k}}^a$ be proportional to  ${\bar{k}}^b$ at points of $\WIH$. 
As a consequence, the existence of commonly used geometrical fields on $\WIH$ and their basic properties --investigated extensively in, e.g. \cite{afk,abl1,akkl1}-- do not use Einstein's or conformal Einstein's equations. Each WIH comes equipped with a degenerate metric $\bq_{ab}$ and a (torsion-free) connection $\bD$ defined intrinsically on $\WIH$, satisfying $\bD_a \bq_{bc} \=0$. Because $\bq_{ab}$ is degenerate, it does \emph{not} determine $\bD$; the connection has extra information. 

As we saw, the condition on the Ricci tensor is satisfied on both:\,
(i) the standard black hole and cosmological WIHs where $(\bM, \bg_{ab})$ is the physical space-time $(M, g_{ab})$ and $\WIH$ is denoted by $\Delta$, and,\, 
(ii) at null infinity with the standard fall-off of matter stress-energy, where now $(\bM, \bg_{ab})$ is the conformal completion $(\hM, \hg_{ab})$ of the physical space-time and $\WIH$ is denoted by $\scrip$.\, 
Therefore both $\Delta$ and $\scri$ are WIHs, and each is equipped with an intrinsically defined degenerate metric and a connection. 
\item Second, in sharp contrast to `stationary horizons' discussed in the literature (e.g. in \cite{cfp}), the intrinsically defined $\bD$ on  WIHs  is generically \emph{time-dependent}! At $\scrip$, the time derivative of this connection  --now denoted by $\hD$-- has direct physical significance: it determines the Bondi news \cite{aa-radiativemodes,aams,ashtekar1987asymptotic}. Consequently, $\hD$ encodes the radiative modes of the gravitational field --two per point of 3-dimensional $\scrip$. Hence there can be flux of gravitational waves carrying BMS-momentum across $\scrip$. 
\item Third, a subtle but key difference arises for {generic} black hole and cosmological WIHs $\Delta$ because they lie in \emph{physical space-times} $(M,\, g_{ab})$: even though the time derivative of the intrinsic connection $D$ is \emph{again non-zero}, there is no flux of gravitational radiation across $\Delta$. We saw that the vacuum Einstein's equations $R_{ab}=0$ imply that the first time derivative of $D$ can be expressed entirely in terms of geometrical fields which are themselves `time independent'. Consequently, the `freely specifiable data' on $\Delta$ consists of fields that can be specified on a 2-dimensional cross section; there are {\emph no volume degrees of freedom} in $D$. Gravitational waves on the other hand correspond to geometric fields that carry 3-dimensional, local degrees of freedom. Because these degrees of freedom are absent, there is no flux of gravitational waves across $\Delta$. Finally, if there are matter fields, then Einstein's equation at $\Delta$ guarantee that no matter flux falls across $\Delta$ either. By contrast, the metric $\h{g}_{ab}$ at $\scrip$ satisfies \emph{conformal} Einstein's equations and they allow the the intrinsic derivative $\hD$ to carry 2 degrees of freedom per point of 3-d $\scrip$, \emph{even though $\h{g}_{ab}$ endows $\scrip$ with a WIH structure.} Note also that it is critical that while $\scrip$ is  a WIH, it is not an IH. Had it been an IH then the connection $\hD$ would have been so constrained that its News tensor $\h{N}_{ab}$ would have been forced to vanish. Thus, the difference between  WIH and IH structures --that may seem like a minor technicality-- makes a profound difference to physics at $\scrip$. 

\end{itemize}

In section \ref{s3} we examined the symmetry groups on WIHs. The group $\G$ on a general $\WIH$ is a 1-dimensional extension of the BMS group by dilations. This extension is essential because, in stationary, non-extremal black hole solutions, for example, the restriction of the Killing field to the horizon is a dilation. The presence of this dilation is directly related to the fact that a generic WIH is naturally equipped with a null normal that is unique only up to a constant rescaling whence a symmetry vector field can rescale this normal by a constant. On the other hand $\scrip$ is an WIH in a conformal completion of the physical space-time and each completion endows $\scrip$ with a specific null normal without the rescaling freedom. Because of this extra structure, the symmetry group $\G$ reduces to the $\B$ group.

In the companion paper \cite{aass2}, we will introduce a new Hamiltonian framework tailored to degrees of freedom that reside in 3-d open regions and apply it to regions $\R$ of $\Delta$ and $\hR$ of $\scrip$ to obtain expressions of fluxes across these regions associated with respective symmetries. Again we will find that while one and the same framework can be applied to both types of WIHs, it leads to strikingly different results in the two cases: At $\scrip$ we obtain the standard BMS fluxes while at $\Delta$ all fluxes vanish. The origin of this difference can be traced back to the fact that at $\scrip$ we have 3-d (radiative) degrees of freedom, while at $\Delta$ we only have 2-d (Coulombic) degrees of freedom. Moreover, in contrast to other approaches, the BMS fluxes (and charges) at $\scrip$ are obtained without having to extend the BMS vector fields $\xi^a$ to the space-time interior. We will compare and contrast the relative merits of this method vis a vis those that have been widely used in the recent literature. Our Hamiltonian framework could be useful also other contexts. For example, it is well-suited to the future space-like infinity in asymptotically de Sitter space-times. 

This unified treatment of black hole (and cosmological) horizons $\Delta$ and null infinity $\scrip$ may open new directions for further research both in classical general relativity and quantum gravity. For example, it was shown in \cite{akkl2} that the expressions of fluxes on \emph{perturbed} black hole and cosmological horizons $\Delta$ are completely analogous to those of perturbations of stationary black holes at $\scrip$. This finding paves the way to  gravitational tomography at late times in compact binary mergers {\cite{aa-banff,aank}, enabling one to read off the late time horizon dynamics from the waveform at $\scrip$, even though the dynamical horizon is causally inaccessible to observers in the asymptotic region. Similarly, in the analysis of isolated gravitating systems \emph{in presence of a positive cosmological constant}, one can use certain cosmological horizons as `local $\scri^\pm$' \cite{Ashtekar_2017,Ashtekar_2019}. The present unified treatment suggests concrete avenues to develop a framework to extract the physics of gravitational waves emitted by these systems. Finally, the availability of a single framework encompassing black hole horizons and null infinity is likely to be useful also in quantum gravity. Specifically, it offers sharper tools to analyze correlations between horizon dynamics and quantum radiation at $\scrip$ during the long semi-classical phase \cite{Ashtekar_2020} of black hole evaporation \cite{aa-loops24}.

\section*{Acknowledgments}

We thank Neev Khera, Maciej Kolanowski, Jerzy Lewandowski and Antoine Rignon-Bret for discussions. This work was supported in part by the Eberly and Atherton research funds of Penn State and the Distinguished Visiting Research Chair program of the Perimeter Institute. 
We thank the referee for a detailed report that led to improvements in the presentation.

\providecommand{\href}[2]{#2}\begingroup\raggedright\endgroup


\end{document}